\documentclass[twocolumn,prb,amsmath,amssymb,floatfix]{revtex4-2} 
\usepackage{amsmath}
\usepackage{amssymb}
\usepackage{graphicx}
\usepackage{bm}
\usepackage{color}
\usepackage{hyperref}
\usepackage{graphicx}
\usepackage{upgreek}
\usepackage{scalerel}
\usepackage{multirow}
 
\usepackage{pgffor}
\usepackage{pdfpages}
\usepackage{mathdots}
\usepackage{float}
\usepackage{silence}
\usepackage{physics} 
\usepackage{lscape}   
\usepackage{color, colortbl}
\usepackage[table]{xcolor}
\usepackage{comment}  
\makeatletter
\AtBeginDocument{\let\LS@rot\@undefined}
\makeatother

\begin{document}

\title{Emergent pair symmetries in systems with poor man's Majorana modes}
\author{Jorge Cayao}
\email[]{jorge.cayao@physics.uu.se}
\affiliation{Department of Physics and Astronomy, Uppsala University, Box 516, S-751 20 Uppsala, Sweden}

\date{\today}
\begin{abstract}
Few-site Kitaev chains are promising for realizing Majorana zero modes without topological protection but fully nonlocal, which are known as poor man's Majorana modes. While several signatures have already been reported both theoretically and experimentally, it still remains unknown what is the nature of superconducting correlations in the presence of poor man's Majorana modes.  In this work, we study few-site Kitaev chains and demonstrate that they host pair correlations with distinct symmetries, entirely determined by the underlying quantum numbers. In particular, we find that a two-site Kitaev chain  hosts local (odd-frequency)  and nonlocal (odd- and even-frequency)  pair correlations, both spin polarized and highly tunable by the system parameters.  Interestingly, the odd-frequency pair correlations exhibit a divergent behaviour around zero frequency  when  the nonlocal $p$-wave pair potential and electron tunneling are of the same order, an effect that can be controlled by the onsite energies. Since a divergent odd-frequency pairing is directly connected to the intrinsic spatial nonlocality of Majorana zero modes in topological superconductors,  the divergent odd-frequency pairing here reflects the intrinsic Majorana nonlocality  of poor man's Majorana modes but without any relation to topology. Our findings could  help understanding the emergent pair correlations in few-site Kitaev chains.  
\end{abstract}

\maketitle

\section{Introduction}
Majorana zero modes (MZMs) emerge in topological superconductors as charge neutral quasiparticles \cite{lutchyn2018majorana,flensberg2021engineered,Marra_2022,tanaka2024theory} and have attracted  an enormous interest  due to their potential for quantum computing applications \cite{Sarma_2015,beenakker2019search,aguado2020majorana,Lahtinen_2017}.   In one dimension,  topological superconductivity with MZMs has been shown to appear in the so-called Kitaev chain \cite{kitaev2001unpaired}, which consists of spin-polarized fermions with $p$-wave superconductivity. Although this type of superconductivity is scarce in nature \cite{RevModPhys.75.657}, it was predicted to occur by combining conventional ingredients such as spin-singlet $s$-wave superconductivity, spin-orbit coupling, and a magnetic field \cite{PhysRevLett.105.077001,PhysRevLett.105.177002}. The simplicity of this proposal has motivated several studies aiming to detect MZMs but, to date, there is no consensus on whether MZMs have been observed or not \cite{prada2019andreev}.

Part of the  challenges is believed to be due to the complex experimental setups \cite{lutchyn2018majorana,flensberg2021engineered}, 
which inevitably enable the presence of other phenomena that obscures Majorana physics \cite{prada2019andreev}. To mitigate some of  the issues,   Kitaev chains with few sites are now being pursued \cite{dvir2023realization,bordin2023crossed,zatelli2023robust,ten_Haaf_2024}, thus offering a bottom-up  engineering approach. In this case, MZMs emerge at  fine-tuned single points in the parameter space, commonly referred to as sweet spots, but do not exhibit  any topological protection \cite{PhysRevB.86.134528}. For this reason, such MZMs were coined   as poor man's Majorana modes (PMMMs); see also Ref.\,\cite{Sau_2012}.  Interestingly, these PMMMs appear as charge neutral quasiparticles, having zero energy and being spatially nonlocal \cite{PhysRevResearch.5.043182,souto2024subgap}, 
in the same way as MZMs in topological superconductors \cite{tanaka2024theory}.   A unique consequence of the charge neutrality of MZMs is that it originates a divergent odd-frequency pairing that is intimately tied to their topology \cite{tanaka2011symmetry,cayao2019odd,mizushima2018multifaceted,tanaka2024theory}, revealing that the superconducting pairing with MZMs is  highly unusual, see also Refs.\,\cite{PhysRevB.87.104513,PhysRevB.92.121404,PhysRevB.95.174516,PhysRevB.95.184506,PhysRevB.96.155426,thanos2019,Takagi18,PhysRevB.101.214507,PhysRevB.101.094506,PhysRevB.106.L100502,ahmed2024x}. 

 Under general conditions, the superconducting pairing can have  even- or odd-frequency symmetries, where   the paired electrons forming Cooper pairs have a  pair amplitude that is \emph{even} or \emph{odd} in their relative time, or frequency, see e. g.\, Refs.\,\cite{Balatsky2017,cayao2019odd}. Identifying the pair symmetries can therefore help understanding the type of emergent superconducting pairing. Since PMMMs exhibit a charge neutrality similar to MZMs but lack of topological protection \cite{PhysRevB.86.134528,PhysRevResearch.5.043182,souto2024subgap}, it is natural to wonder what is the nature of the emergent superconducting pairing in the presence of PMMMs.

 \begin{figure}[!t]
\centering
	\includegraphics[width=0.47\textwidth]{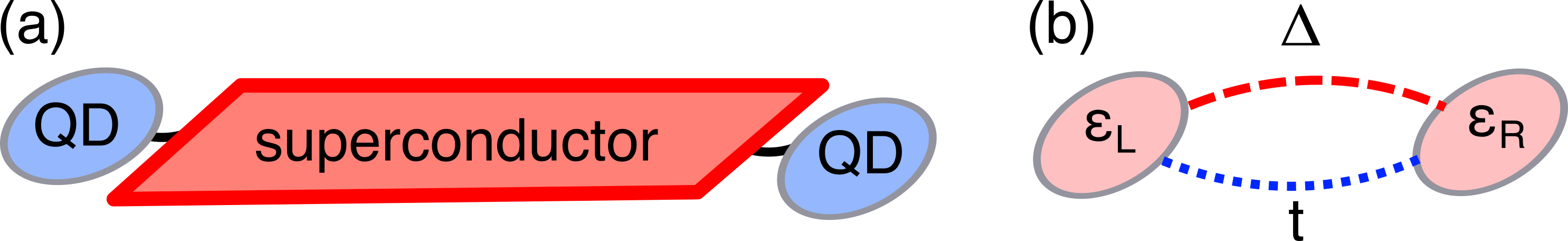}
 	\caption{(a) A superconductor with Rashba spin-orbit coupling (red) under a magnetic field is coupled to two quantum dots (QDs in blue). (b) The QDs become superconducting (light red) with a spin-polarized $p$-wave pair potential $\Delta$  and a hopping amplitude $t$, effectively realizing a two-site Kitaev chain. The onsite energies of the QDs is denoted by $\varepsilon_{\rm L,R}$. }
\label{Fig1} 
\end{figure}

In this work, we consider a few-site Kitaev chain [Fig.\,\ref{Fig1}] and investigate the emergence of superconducting pair correlations. By carrying out a full symmetry classification that involves  all the quantum numbers  in these type of systems, we find that there are multiple pair symmetries naturally emerging both locally and nonlocally. In general, local superconducting pair amplitudes  appear with an odd-frequency dependence, while nonlocally the pair amplitudes can exhibit even- and odd-frequency profiles. More specifically, we demonstrate in a two-site Kitaev chain that  the local pair correlations develop an odd-frequency dependence  with a divergent  profile around zero frequency when   the nonlocal $p$-wave pair potential and  electron tunneling are equal (sweet spot) and at least one onsite energy vanishes. Furthermore, the nonlocal pair correlations also exhibit a similar divergent frequency profile but, unlike the local pairing, requires having distinct onsite energies and at least one of them vanishing.     Away from the stringent sweet spot conditions, the local and nonlocal odd-frequency pair amplitudes follow a linear frequency dependence, reaching zero value at zero frequency.  The divergent odd-frequency pairing reveals a   behavior of the emergent superconducting correlations that is similar to what occurs in topological superconductors with MZMs but here without any relation to topology.
Our findings might be useful for understanding the type of emergent superconducting correlations in few-site Kitaev chains with PMMMs.

 The remainder of this work is organized as follows. In Sec.~\ref{section2} we discuss   the allowed pair symmetries in few-site Kitaev chains. In   Sec.~\ref{section3} we present the model of a two-site Kitaev chain, discuss its spectral properties, and  demonstrate the emergence of local and nonlocal superconducting pair correlations.  Finally,  we present our conclusions in   Sec.~\ref{section4},

\section{Pair symmetries in few-site Kitaev chains}
\label{section2}
We begin by providing a general symmetry classification of the superconducting pair correlations  allowed by all the  quantum numbers in few-site Kitaev chains. 
To characterize the superconducting pair correlations, we employ the anomalous Green's function defined as   ${\mathcal F}^{nm}_{\alpha\beta}(t,t')=\langle \mathcal{T}c_{\alpha n}(t)c_{\beta m}(t')\rangle$ where $\mathcal{T}$ is the time ordering operator, $c_{n\alpha}(t)$ annihilates an electronic state in site $\alpha$ of superconductor $n$ at time $t$ \cite{mahan2013many,zagoskin}. 
The anomalous Green's function ${\mathcal F}^{nm}_{\alpha\beta}(t,t')$ is also known as pair amplitude or pair correlation and  its formation is fully tied to the quantum numbers of the two paired electrons, namely,  $(n,m)$, $(\alpha,\beta)$, and    time coordinates ($t,t'$). For instance, the spin configuration of the Kitaev chain already dictates the spin symmetry of ${\mathcal F}^{nm}_{\alpha\beta}$ since we do not assume any active spin field: the Kitaev chain consists of spin-polarized fermions \cite{kitaev2001unpaired}, which implies that all the fermions  have the same spin and that the pair amplitude  has a spin-triplet symmetry \cite{PhysRevB.87.104513,PhysRevB.92.121404,PhysRevB.95.174516,PhysRevB.95.184506,thanos2019,Takagi18,PhysRevB.101.214507,PhysRevB.101.094506}; see also Ref.\,\cite{RevModPhys.63.239}. 

The symmetries with respect to the other quantum numbers $(n,m)$, $(\alpha,\beta)$, and    time coordinates ($t,t'$) are not arbitrary but must respect the fermionic nature of the composed electrons. Since the pair amplitude  behaves as a two-electron wavefunction, it must fulfill the antisymmetry condition such that ${\mathcal F}^{nm}_{\alpha\beta}(t,t')=-{\mathcal F}^{mn}_{\beta\alpha}(t',t)$ under the total exchange of quantum numbers plus exchange of time coordinates \cite{tanaka2011symmetry,cayao2019odd,triola2020role}. Similarly, in frequency domain, the antisymmetry condition implies  ${F}^{nm}_{\alpha\beta}(\omega)=-{F}^{mn}_{\beta\alpha}(-\omega)$, where ${F}^{nm}_{\alpha\beta}(\omega)$ is the Fourier transform of ${\mathcal F}^{nm}_{\alpha\beta}(t,t')$.  Thus,  the allowed pair  symmetries must be antisymmetric under a total exchange of quantum numbers.  Taking this into account, under the individual exchange of frequency ($\omega$), site indices $(\alpha, \beta)$, and superconductor indices $(n, m)$, the pair amplitude can be either an \emph{even} (E) or \emph{odd} (O) function.  As pointed out in the previous paragraph,  the Kitaev chain involves spin-polarized fermions which leaves the spin symmetry to be triplet (T).    We find that there are four distinct spin triplet pair symmetries that obey the antisymmetry condition, see Tab.\, \ref{table:1}.   It can be seen that the  superconductor index $n$ (sup. index), as well as the dot index $\alpha$, play a  role that is similar to the band index in multiband superconductors  \cite{PhysRevB.88.104514,PhysRevB.92.224508,PhysRevB.101.184501,PhysRevB.102.184506,triola2020role} and sideband index in Floquet superconductors \cite{PhysRevB.103.104505,PhysRevB.109.134517}, see also Ref.\,\footnote{It is worth noting that the role of the dot index for broadening the emerging superconducting symmetries was studied before, see e.g.\,Refs.\,\cite{PhysRevB.90.220501,PhysRevB.93.201402,PhysRevB.109.205406}. However, none of these previous works addressed   few-site Kitaev chains.}.   The sup. index $n$ becomes active when multiple few-site Kitaev chains are coupled, e.g., in Josephson junctions \cite{PhysRevB.109.075101}, which are expected to be useful for realizing superconducting circuits.
However, by focusing on the pair amplitudes inside a given superconductor, the sup. index restricts its symmetry to be \emph{even} and only two pair symmetry classes are allowed:   even-frequency, spin-triplet, odd under dot index, even under sup. index or ETOE; odd-frequency, spin-triplet, even under dot index, even under sup. index or OTEE.  While in general all the allowed pair symmetries can appear as a linear combination, and can therefore coexist, the dominance of a particular symmetry characterizes the type of emergent superconducting pairing. Therefore, under general circumstances,  few-site Kitaev chains are expected to host distinct types of pair symmetry classes.

\begin{table}[!t]
\centering
\begin{tabular}{ |c|c|c|c||c|  }
 \hline
  \hline
Frequency & Spin & Site index & Sup. index &Class\\
  ($\omega\leftrightarrow -\omega$)& ($\uparrow \leftrightarrow \downarrow$) & ($\alpha\leftrightarrow \beta$)& ($n\leftrightarrow m$)& (total exchange) \\ [0.5ex] 
 \hline
  \hline
 {\bf E}ven &{\bf T}riplet &  {\bf E}ven&  {\bf O}dd&ETEO\\
  {\bf E}ven &{\bf T}riplet &  {\bf O}dd&  {\bf E}ven&ETOE\\
{\bf O}dd &{\bf T}riplet &  {\bf E}ven&  {\bf E}ven&OTEE\\
{\bf O}dd&{\bf T}riplet &  {\bf O}dd&  {\bf O}dd&OTOO\\
 \hline
\end{tabular}
\caption{Allowed superconducting pair symmetries in few-site Kitaev chains as a result of the total antisymmetrization of the pair amplitudes when exchanging frequency, spins, site index, and superconducting (sup.) index. In a two-site Kitaev chain, which involves a single superconducting system, only ETOE and OTEE classes are present.}
\label{table:1}
\end{table}

\section{Emergent pair correlations in two-site Kitaev chains}
\label{section3}
Having discussed  the allowed pair symmetries in few-site Kitaev chains, now we explore their formation in a two-site Kitaev chain which consists of two spin-polarized coupled sites  with $p$-wave pairing, see Fig.\,\ref{Fig1}(b).  In the basis $\Psi=(c_{\rm L},c_{\rm R},c_{\rm L}^{\dagger},c_{\rm R}^{\dagger})$, the two-site Kitaev chain is  modelled by the following Bogoliubov-de Gennes (BdG) Hamiltonian,
\begin{equation}
\label{Eq1}
H_{\rm BdG}
= \varepsilon_{\rm L}\eta_{+}\tau_{z}+\varepsilon_{\rm R}\eta_{-}\tau_{z}+t\eta_{x}\tau_{z}-\Delta\eta_{y}\tau_{y}\,,
 \end{equation}
where $\varepsilon_{\alpha}$ is the onsite energy of dot $\alpha=L/R$, $t$ is the hopping amplitude, and $\Delta$ the $p$-wave pair potential. Moreover,  $\eta_{\pm}=(\eta_{0}\pm\eta_{z})/2$, with  $\eta_{i}$    the $i$-th Pauli matrix in the site subspace, while $\tau_{i}$ the Pauli matrix in   Nambu  subspace. Notably, the minimal Kitaev chain given by Eq.\,(\ref{Eq1}) at the sweet spot $\varepsilon_{\rm L,R}=0$ and $\Delta=t$ hosts a pair of MZMs, with their wavefunctions  fully  located at the left and right sites but without being topologically protected, namely, Eq.\,(\ref{Eq1}) hosts a pair of  PMMMs \cite{PhysRevB.86.134528}. Moreover, it is worth noting that two-site Kitaev chain also holds experimental relevance as it has been recently realized in superconductor-semiconductor hybrids \cite{dvir2023realization}, see Fig.\,\ref{Fig1}(a). Most of the properties of PMMMs have been shown to be similar to those of MZMs in topological superconductors. However, the nature of the induce superconducting pairing is still unknown. In particular,  being the PMMMs of Majorana origin and having an intrinsic charge neutrality poses the question about the type of induced superconducting pairing.

We are here interested in the emergent pair symmetries under the presence of PMMMs, which, as discussed in Sec.\,\ref{section3}, requires the calculation of the anomalous electron-hole Green's function. For this purpose, we obtain the Green's function of the BdG Hamiltonian given by Eq.\,(\ref{Eq1})
\begin{equation}
\label{GF}
\mathcal{G}(\omega)=
\begin{pmatrix}
G(\omega)&F(\omega)\\
\bar{F}(\omega)&\bar{G}(\omega)
\end{pmatrix}=(\omega-H_{\rm BdG})^{-1}\,,
\end{equation}  
where $\omega$ represents complex frequencies unless otherwise specified. Here, the diagonal components ($G$ and $\bar{G}$) represent the normal electron-electron and hole-hole Green's functions, while ($F$ and $\bar{F}$) are the anomalous electron-hole and hole-electron Green's functions. Note that the normal and anomalous Green's functions are still matrices in the subspace spanned by the two sites, see also Eq.\,(\ref{Eq1}).  While $G$ enables the calculation of the spectral function, $F$ determines the superconducting pairing. For completeness, in what follows we explore first  the spectral function and then focus on the emergent pair symmetries.
 
\subsection{Majorana signatures in the spectral function}
\label{section4a}
\begin{figure}[!t]
\centering
	\includegraphics[width=0.49\textwidth]{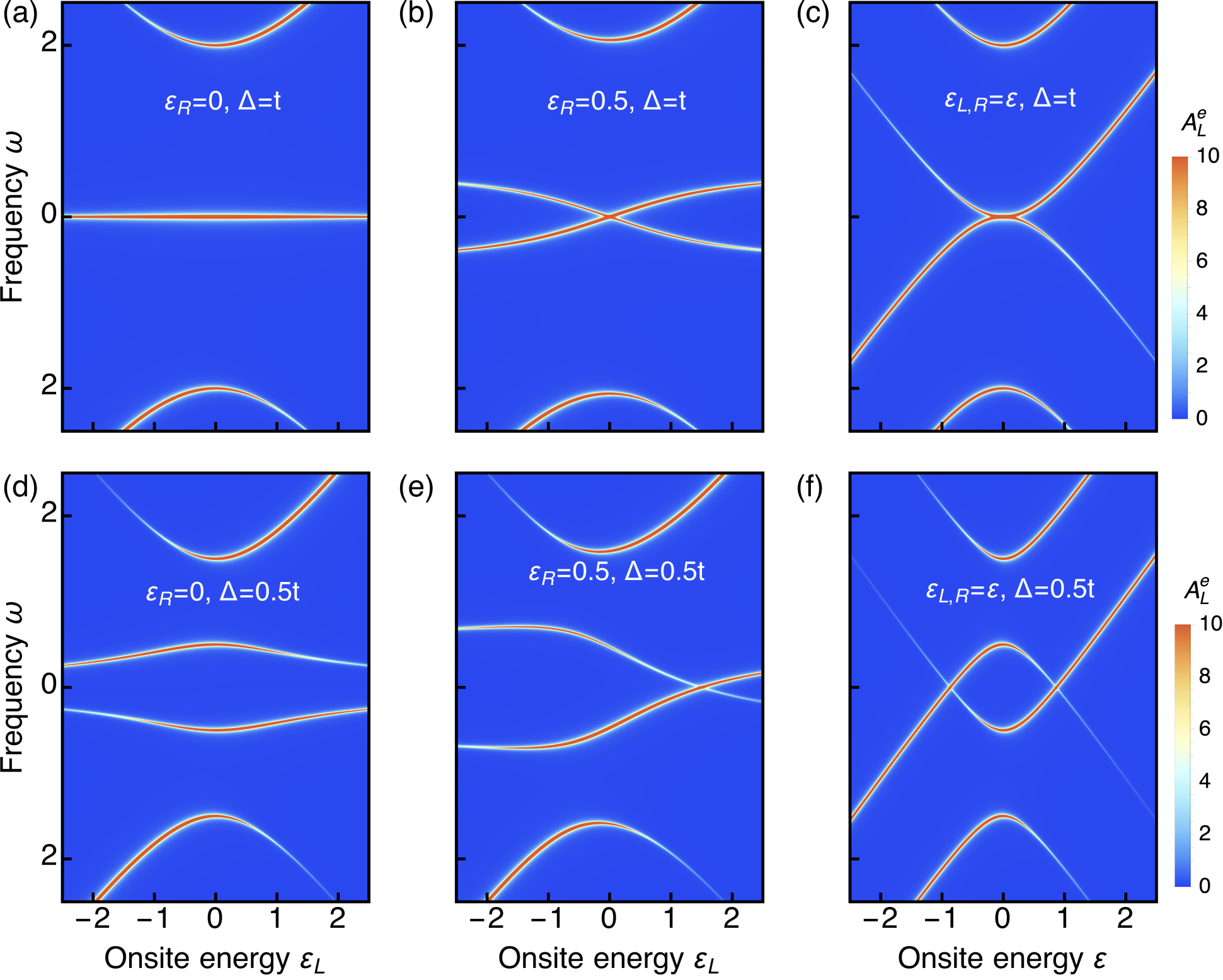}
 	\caption{Spectral function in the left dot $A_{\rm L}^{e}$ as a function of frequency $\omega$  and onsite energies. Panels (a-c) correspond to $\Delta=t$ with $\varepsilon_{\rm R}=0$, $\varepsilon_{\rm R}=0.5$, $\varepsilon_{\rm L,R}=\varepsilon$. Panels (d-f) the same as in (a-c) but at $\Delta=0.5t$.}
\label{Fig2} 
\end{figure}

Before addressing the pair correlations, it is worth discussing  the formation of PMMMs. While this can be carried out in different ways, it is useful to explore the spectral function not only because it can be directly obtained from the normal Green's functions obtained above but, importantly, also because it can be experimentally accessed via conductance measurements \cite{datta1997electronic}. Since conductance is often measured in Majorana experiments \cite{prada2019andreev,dvir2023realization}, the spectral signatures of PMMMs can help understanding their formation.  By using Eq.\,(\ref{Eq1}) and Eq.\,(\ref{GF}), we find  the normal Green's function components given by
\begin{equation}
\label{Gnormal}
\begin{split}
G_{\rm LL}(\omega)&=
\frac{t^{2}(\varepsilon_{\rm R}-\omega)-(\varepsilon_{\rm R}+\omega)
P_{\rm LR}(\omega)}{D(\omega)}\,,\\
G_{\rm RR}(\omega)&=\frac{t^{2}(\varepsilon_{\rm L}-\omega)-(\varepsilon_{\rm L}+\omega)
P_{\rm RL}(\omega)}{D(\omega)}\,,\\
G_{\rm LR}(\omega)&=\frac{t[\Delta^{2}-t^{2}+(\varepsilon_{\rm L}+\omega)(\varepsilon_{\rm R}+\omega)]}{D(\omega)}\,,\\
G_{\rm RL}(\omega)&=G_{\rm LR}(\omega)\,,
\end{split}
\end{equation}
where $P_{\rm LR(RL)}(\omega)=[\Delta^{2}+(\varepsilon_{\rm L(R)}+\omega)(\varepsilon_{\rm R(L)}-\omega)]$ and $D(\omega)=(\Delta^{2}-t^{2}+\varepsilon_{\rm L}\varepsilon_{\rm R})^{2}-(2t^{2}+2\Delta^{2}+\varepsilon_{\rm L}^{2}+\varepsilon_{\rm R}^{2})\omega^{2}+\omega^{4}$. Also, $\bar{G}_{\rm LL(RR)}=G_{\rm LL(RR)}(\varepsilon_{\alpha}\rightarrow-\varepsilon_{\alpha})$, $\bar{G}_{\rm LR(RL)}=-G_{\rm LR(RL)}(\varepsilon_{\alpha}\rightarrow-\varepsilon_{\alpha})$. For simplicity but without loss of generality we focus on the spectral function in the left site obtained as $A_{\rm L}^{e}(\omega)=-{\rm Im Tr}G_{\rm LL}(\omega+i\eta)$ where $\eta$ is an  infinitesimal positive number enabling the   analytic continuation to real frequencies $\omega$ \cite{mahan2013many}.  To access $A_{\rm L}^{e}$ via conductance, the left site of the two-site Kitaev chain can be coupled to a normal lead \cite{cayao2024NHPMMM,datta1997electronic}.  In Fig.\,\ref{Fig2} we present $A_{\rm L}^{e}(\omega)$ as a function of real frequency $\omega$ and onsite energies $\varepsilon_{\rm L,R}$. The top and bottom rows correspond to the spectral function at sweet spot $\Delta=t$ and away from it $\Delta\neq t$. The immediate feature we observe is that  the spectral function reveals the energy levels of the two-site Kitaev chain \cite{PhysRevB.86.134528}  which read $E^{s}_{\pm}=\pm|\sqrt{\smash[b]{t^{2}+\varepsilon_{-}^{2}}}-(-1)^{s} \sqrt{\smash[b]{\Delta^{2}+\varepsilon_{+}^{2}}}|$, where  $\varepsilon_{\pm}=(\varepsilon_{\rm L}\pm \varepsilon_{\rm R})/2$ and  $s=0$ ($s=1$) labels the two lowest (excited) levels. The features of the energy levels in spectral function is of course expected because the poles of the Green's function $G_{\rm LL}$ are directly connected to the spectrum of $H_{\rm BdG}$, see Eq.\,(\ref{GF}).

In the sweet spot $\Delta=t$, when one of the onsite energies vanishes, the spectral function $A_{\rm L}^{e}(\omega)$ develops a large value at zero frequency, as seen in Fig.\,\ref{Fig2}(a). This can be understood by noting that $G_{\rm LL}(\omega)$ given in Eqs.\,(\ref{Gnormal}) develops a zero-frequency resonance at $\varepsilon_{\rm R}=0$ and $\Delta=t$, namely,  $G_{\rm LL}(\omega)=[\omega (\omega+\varepsilon_{\rm L})-2\Delta^{2}]/\{\omega[\omega^{2}-(4\Delta^{2}+\varepsilon_{\rm L}^{2})]\}$; the zero-frequency resonance is then evident from the $1/\omega$ dependence of $G_{\rm LL}$. This zero frequency resonance corresponds to having $E^{0}_{\pm}=0$ at $\Delta=t$ and $\varepsilon_{\rm R}=0$, which corresponds to the energy of PMMMs \cite{PhysRevB.86.134528}.  By inducing a finite onsite energy in the right dot, the spectral function develops a bowtie-like profile around zero frequency and acquires large zero-frequency values  only when the left onsite energy vanishes, as seen in Fig.\,\ref{Fig2}(b). Similarly, in the case of having equal onsite energies, the spectral function  becomes large at zero frequency only when such energies vanish, see Fig.\,\ref{Fig2}(c). We can therefore conclude that the emergence of PMMMs is reflected in large zero-frequency values of the spectral function.

Away from the sweet spot condition, when $\Delta\neq t$, the situation becomes drastically different, see Fig.\,\ref{Fig2}(d-f). In this case, when one of the onsite energies is fixed at zero, the spectral function acquires a diamond-like profile around zero frequency but does not reach large zero-frequency values, see Fig.\,\ref{Fig2}(d). A finite right onsite energy  then leads to an asymmetric profile with respect $\varepsilon_{\rm R}=0$, inducing a large zero-frequency spectral weight at $\varepsilon_{\rm L}=(t^{2}-\Delta^{2})/\varepsilon_{\rm R}$. For equal onsite energies $\varepsilon_{\rm L,R}=\varepsilon$, the spectral function exhibits large zero frequency values at $\varepsilon=\pm\sqrt{t^{2}-\Delta^{2}}$, as observed in Fig.\,\ref{Fig2}(f). Away from the sweet spot $\Delta\neq t$, however, no PMMMs exist \cite{PhysRevB.86.134528}. Only at the sweet spot stable zero energy states appear with large spectral weights.

\subsection{Emergent pair amplitudes}
\label{section4b}
We now explore the symmetries of the emergent pair amplitudes, which we obtain from the anomalous Green's functions. By using Eq.\,(\ref{Eq1}) and Eq.\,(\ref{GF}), we find that the anomalous Green's function components are given by
\begin{equation}
\label{Fanomal}
\begin{split}
F_{\rm LL}(\omega)&=\frac{2\omega t\Delta}{D(\omega)}\,,\\
F_{\rm RR}(\omega)&=-\frac{2\omega t\Delta}{D(\omega)}\,,\\
F_{\rm LR}^{+}(\omega)&=\frac{\omega(\varepsilon_{\rm R}-\varepsilon_{\rm L})\Delta}{D(\omega)}\,,\\
F_{\rm LR}^{-}(\omega)&=\frac{\Delta(\Delta^{2}-t^{2}+\varepsilon_{\rm L}\varepsilon_{\rm R}-\omega^{2})}{D(\omega)}\,,
\end{split}
\end{equation}
where $\omega$ represents complex frequencies, $F_{\rm LR}^{\pm}=(F_{\rm LR}\pm F_{\rm RL})/2$, and $D(\omega)$ is given below Eq.\,(\ref{Gnormal}).  Moreover, $\bar{F}_{\rm LL(RR)}=F_{\rm LL(RR)}(\varepsilon_{\alpha}\rightarrow-\varepsilon_{\alpha})$ and  $\bar{F}_{\rm LR(RL)}^{\pm}=-F^{\pm}_{\rm LR(RL)}(\varepsilon_{\alpha}\rightarrow-\varepsilon_{\alpha})$.  Eqs.\,(\ref{Fanomal}) represent the emergent superconducting pair amplitudes in a two-site Kitaev chain, which puts us in position  to identify their symmetries following Sec.\,\ref{section3}. As discussed in Sec.\,\ref{section3}, the pair symmetries depend on the quantum numbers associated to the site indices, sup. indices, spins, and frequency.  Since  here we consider a single superconducting system, the pair amplitudes can only be \textit{even}   under the exchange of such a sup. index. Moreover, since we deal with spinless fermions, the pair amplitudes have a \textit{spin-triplet}  symmetry, see Sec.\,\ref{section3}. Thus, all the pair amplitudes in Eqs.\,(\ref{Fanomal}) are \textit{spin-triplet} and \textit{even} in the sup. index. The remaining symmetries, with respect to the site indices and frequency, however, are not the same for all the pair amplitudes in Eqs.\,(\ref{Fanomal}) and, as we will see below, will determine the type of emergent superconducting pairings.

Locally, the onsite pair amplitudes $F_{\rm LL(RR)}$ are evidently \textit{even} under the exchange of the site index $\alpha=L,R$ and \textit{odd}  functions of   frequency $\omega$; note that $D(\omega)=D(-\omega)$ is an even function of $\omega$ and  $F_{\rm LL(RR)}$ exhibit opposite sign \footnote{The opposite signs of the local pair amplitudes ($F_{\rm LL}$ and $F_{\rm RR}$) have also been reported  in the local pair amplitudes  of finite topological superconductors with MZMs \cite{tanaka2024theory}; in our case, however, no topology is involved. Moreover, the  Majorana polarization  has also been shown to develop opposite signs at the edges  of   topological superconductors  \cite{awoga2024imajor}. This suggests an intriguing relation between odd-frequency pairing and Majorana polarization.}. Thus, the local pair amplitudes exhibit an odd-frequency, spin-triplet, even-site, even-sup. symmetry, which  corresponds to the symmetry class OTEE of Tab.\,\ref{table:1}. This pair symmetry class is the only type of local emergent superconducting pairing emerging in these type of systems;   note that the two-site Kitaev chain does not have a  local superconducting pair potential, revealing that the OTEE class is indeed an emergent superconducting pairing. When it comes to the nonlocal pair amplitudes $F^{\pm}_{\rm LR}$, the expressions given in  Eqs.\,(\ref{Fanomal}) are already symmetrized with respect to the site index: $F^{+}_{\rm LR}$ is  \textit{even} under the exchange of L and R, while  $F^{-}_{\rm LR}$ is  \textit{odd}. Moreover, Eqs.\,(\ref{Fanomal}) already reveal that $F^{+}_{\rm LR}$ and  is an \textit{odd} function of $\omega$ while $F^{-}_{\rm LR}$ is an \textit{even} function of $\omega$. With these considerations, together with the \textit{spin-triplet} and \textit{even}-sup. index symmetries, it is already clear that $F^{+}_{\rm LR}$ and $F^{-}_{\rm LR}$ correspond to the OTEE and ETOE symmetry classes of Tab.\,\ref{table:1}. Thus, nonlocally, the two-site Kitaev chain hosts two types of superconducting correlations, with the OTEE being entirely induced while the ETOE pairing tied to the parent $p$-wave pair potential.

Therefore, two-site Kitaev chains host two types of superconducting pairing, OTEE locally while OTEE and ETOE nonlocally, provided the parent superconductor is  spin-polarized and $p$-wave. It is now necessary to   inspect the behavior of these emergent pair correlations as a function of the system parameters and contrast their presence under the presence and absence of PMMMs.

\subsection{Pair amplitudes at the sweet spot $\Delta=t$}
\label{section4c}
\begin{figure}[!t]
\centering
	\includegraphics[width=0.49\textwidth]{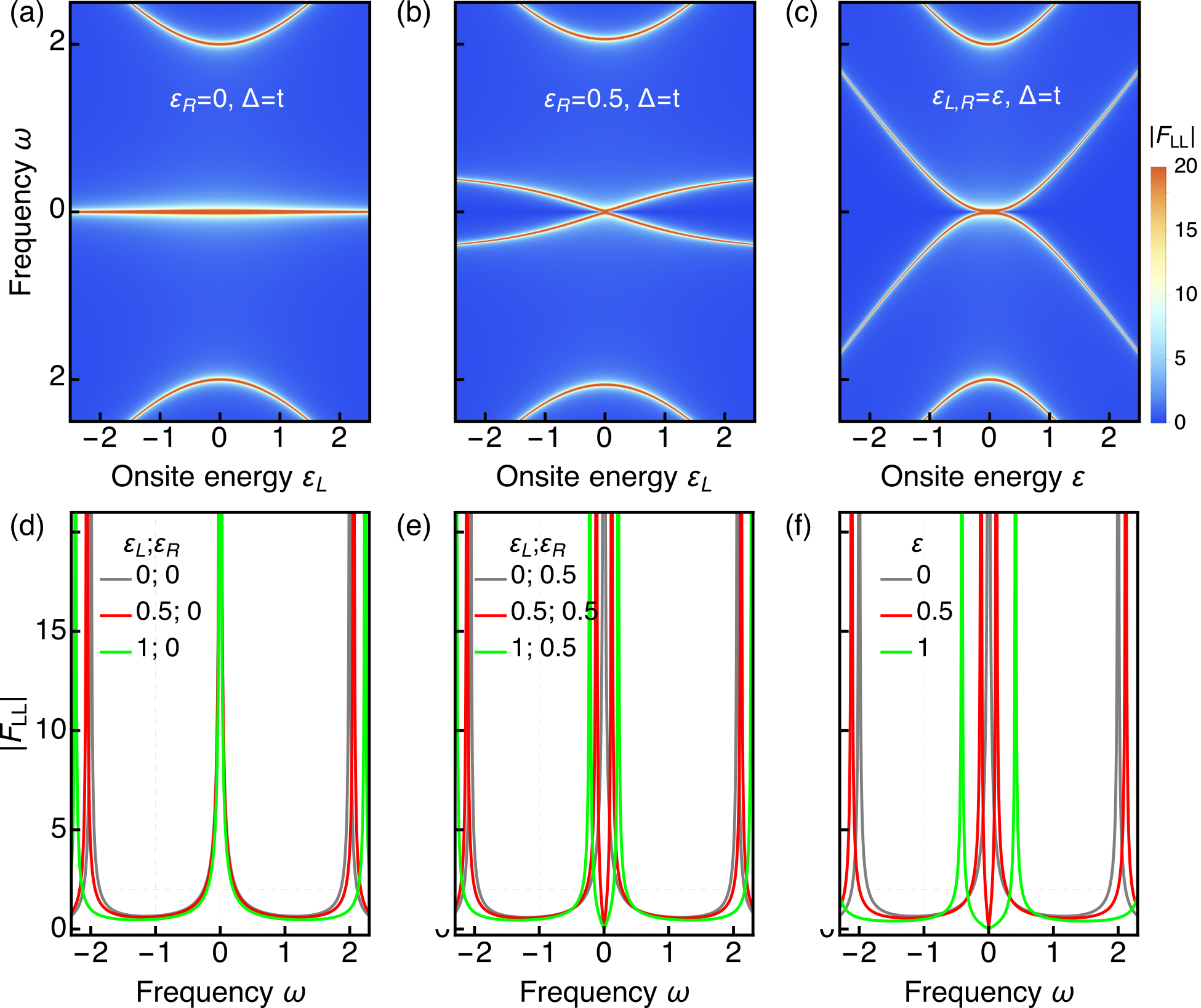}
 	\caption{(a-c) Absolute value of the local pair amplitude in the left dot $|F_{\rm LL}|$ 
as a function of frequency $\omega$ and onsite energy $\varepsilon_{\rm L}$   at $\Delta=t$ and $\varepsilon_{\rm R}=0$, $0.5$, $\varepsilon_{\rm L}$. (d-f) $|F_{\rm LL}|$ as a function of $\omega$ at distinct $\varepsilon_{\rm L,R}$. }
\label{Fig3} 
\end{figure}

\begin{figure*}[!t]
\centering
	\includegraphics[width=0.95\textwidth]{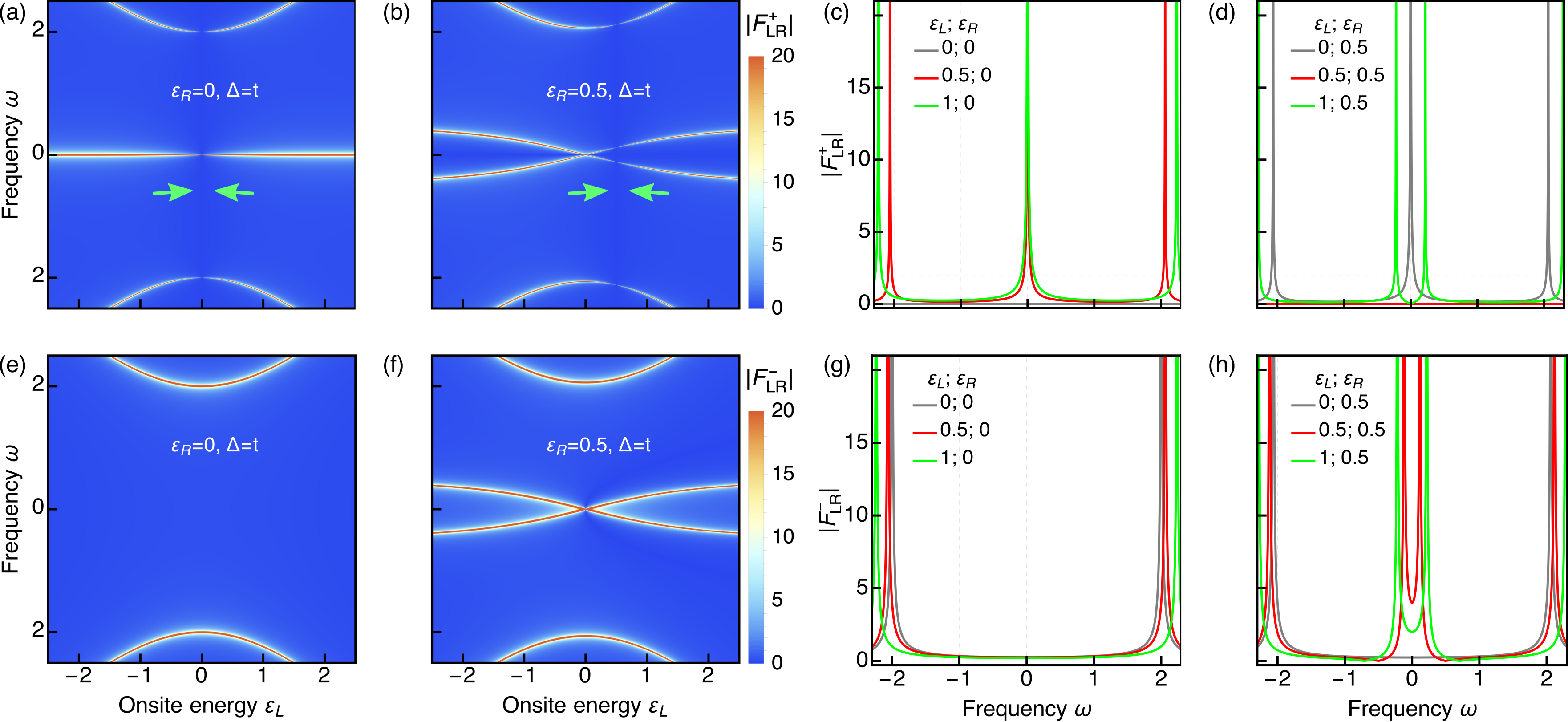}
 	\caption{(a,b) Absolute value of the nonlocal pair amplitudes $|F_{\rm LR}^{+}|$ as a function of frequency $\omega$ and onsite energy $\varepsilon_{\rm L}$   at $\Delta=t$ and $\varepsilon_{\rm R}=0$, $0.5$. (c,d)  $|F_{\rm LR}^{+}|$ as a function of $\omega$ at distinct values of $\varepsilon_{\rm L,R}$. (e,f,g,h) Same as   (a,b,c,d) but for $|F_{\rm LR}^{-}|$. }
\label{Fig5} 
\end{figure*}

In the sweet spot regime $\Delta=t$, but still maintaining finite onsite energies, the pair amplitudes given by Eqs.\,(\ref{Fanomal}) are given by 
\begin{equation}
\label{FanoSS}
\begin{split}
F_{\rm LL}(\omega)&=\frac{2\omega t^{2}}{
\omega^{4}
-(4t^{2}+\varepsilon_{\rm L}^{2}+\varepsilon_{\rm R}^{2})\omega^{2}+\varepsilon_{\rm L}^{2}\varepsilon_{\rm R}^{2}}\,,\\
F_{\rm RR}(\omega)&=-\frac{2\omega t^{2}}{
\omega^{4}
-(4t^{2}+\varepsilon_{\rm L}^{2}+\varepsilon_{\rm R}^{2})\omega^{2}+\varepsilon_{\rm L}^{2}\varepsilon_{\rm R}^{2}}\,,\\
F^{+}_{\rm LR}(\omega)&=\frac{\omega(\varepsilon_{\rm R}-\varepsilon_{\rm L})t}{
\omega^{4}
-(4t^{2}+\varepsilon_{\rm L}^{2}+\varepsilon_{\rm R}^{2})\omega^{2}+\varepsilon_{\rm L}^{2}\varepsilon_{\rm R}^{2}}\,,\\
F^{-}_{\rm RL}(\omega)&=\frac{(\varepsilon_{\rm L}\varepsilon_{\rm R}-\omega^{2})t}{
\omega^{4}
-(4t^{2}+\varepsilon_{\rm L}^{2}+\varepsilon_{\rm R}^{2})\omega^{2}+\varepsilon_{\rm L}^{2}\varepsilon_{\rm R}^{2}}\,.
\end{split}
\end{equation}
To visualize the behaviour of these pair amplitudes, in Fig.\,\ref{Fig3} and Fig.\,\ref{Fig5} we plot the absolute value of the local  ($F_{\rm LL}$) and nonlocal ($F_{\rm LR}^{\pm}$)  amplitudes in the sweet spot $\Delta=t$. In both cases we show the pair amplitudes as a function of   frequency $\omega$ and onsite energies $\varepsilon_{\rm L,R}$ [Fig.\,\ref{Fig3}(a-c) and Fig.\,\ref{Fig5}(a,b,e,f)] and also the sole frequency dependence at fixed  $\varepsilon_{\rm L,R}$  [Fig.\,\ref{Fig3}(d-f) and Fig.\,\ref{Fig5}(c,d,g,h)]. The first feature we identify is that the pair amplitudes reveal the formation of the energy levels discussed in the spectral function in the previous subsection, expected because both quantities share the same denominator, see   Eqs.\,(\ref{Fanomal})  (and also Eqs.\,(\ref{FanoSS}) and Eqs.\,(\ref{Gnormal}). The numerators of Eqs.\,(\ref{Fanomal}) develop some interesting dependences purely associated to the type of emergent pair correlation.

In the case of the local pair amplitudes $F_{\rm LL/RR}$, when one of the onsite energy vanishes (e.g., $\varepsilon_{\rm R}=0$), we find that they acquire large zero-frequency values irrespective of the finite value of 
the other onsite energy (e.g., $\varepsilon_{\rm L}$), see Fig.\,\ref{Fig3}(a) for $|F_{\rm LL}|$.  By inspecting the frequency dependence of  $|F_{\rm LL}|$ at $\varepsilon_{\rm R}=0$ and  different  $\varepsilon_{\rm L}$ in Fig.\,\ref{Fig3}(d), we observe that $|F_{\rm LL}|$ has a divergent profile around $\omega=0$ irrespective of the value of  $\varepsilon_{\rm L}$; see also gray curves in Fig.\,\ref{Fig3}(e,f).  To understand this intriguing divergent frequency dependence, we write down the local pair amplitudes given by Eqs.\,(\ref{FanoSS}) at $\varepsilon_{\rm R}=0$, obtaining
\begin{equation}
\label{EQSS}
\begin{split}
F_{\rm LL}(\omega)&=\frac{1}{\omega}
\left[\frac{2 t^{2}}{
\omega^{2}
-(4t^{2}+\varepsilon_{\rm L}^{2})}\right]\,,\\
F_{\rm RR}(\omega)&=-\frac{1}{\omega}
\left[
\frac{2 t^{2}}{
\omega^{2}-(4t^{2}+\varepsilon_{\rm L}^{2})}\right]\,,
\end{split}
\end{equation}
which at low frequencies can be approximated by  $F_{\rm LL/RR}(\omega)\approx \mp [1/(2\omega)]\mp [\omega/(8\Delta)]$. It is thus evident the emergence of local pair amplitudes with a divergent   profile near zero frequency [Fig.\,\ref{Fig3}(a,d)]. Since the local pair amplitudes correspond to the OTEE pair symmetry class, their odd-frequency symmetry is fully determined by having a divergent frequency dependence. As noted in the introduction, having divergent odd-frequency pair amplitudes is a unique property of   topological superconductors with MZMs, associated to topology and revealing the   Majorana  nonlocality \cite{PhysRevB.87.104513,PhysRevB.92.121404,PhysRevB.95.174516,PhysRevB.95.184506,thanos2019,Takagi18,PhysRevB.101.214507,PhysRevB.101.094506}; see also Refs.\,\cite{tanaka2011symmetry,cayao2019odd,mizushima2018multifaceted,tanaka2024theory}. 
 In the present case, however, our minimal two-site Kitaev chain in the sweet spot hosts PMMMs without any topological properties but, surprisingly, we find divergent odd-frequency pairing. This occurs because even in the two-site Kitaev chain, it is possible to realize a pair of fully nonlocal zero-energy PMMMs in the same way as MZMs.  The relation between Majorana nonlocality and divergent odd-frequency pairing can be further understood by noting that   Majorana operators   fulfil  a self-conjugation property  $\gamma_{i}^{\dagger}=\gamma_{i}$, which occurs when  MZMs are fully nonlocal and charge neutral \cite{tanaka2024theory},  namely, well separated from each  other without any energy splitting and spatial overlap between them. Thus, taking into account   $\gamma_{i}^{\dagger}=\gamma_{i}$, the anomalous pair correlation ($f$) of a Majorana operator can be written as    $f\sim\langle\gamma_{i}\gamma_{i}^{\dagger} \rangle=\langle\gamma_{i} \gamma_{i} \rangle=1/\omega$. Thus, the Majorana nonlocality, inherited  by both MZMs in topological superconductors and PMMMs in few-site Kitaev chains, naturally produce a divergent odd-frequency pairing as a strong signature of their unusual superconducting pairing.
 
The large values of $|F_{\rm LL}|$ are also seen when $\varepsilon_{\rm R}\neq0$ in Fig.\,\ref{Fig3}(b,e),  where the local pair amplitude forms a divergent profile near zero frequency only when $\varepsilon_{\rm L}=0$, consistent with the discussion presented in the previous paragraph. A similar behavior occurs when both onsite energies are equal in   Fig.\,\ref{Fig3}(c,f), giving rise to divergent values of $|F_{\rm LL}|$ only when both energies vanish. Thus, having both onsite energies with finite values results in local pair amplitudes that vanish at $\omega=0$, which occurs because in this case there is a linear frequency dependence in the numerator of Eqs.\,(\ref{FanoSS}); hence  the local pair amplitudes approach linearly to zero frequency. This behaviour is observed in red and green curves of Fig.\,\ref{Fig3}(e,f), supporting the idea that the divergent frequency behaviour is a unique property of the sweet spot regime with $\Delta=t$ and vanishing values of either of the onsite energies. 

For the nonlocal pair amplitudes  $F^{\pm}_{\rm LR}$, both correspond to distinct pair symmetries and this is reflected in Fig.\,\ref{Fig5}(a-d) and Fig.\,\ref{Fig5}(e-h) for the OTEE    $F^{+}_{\rm LR}$ and ETOE   pair symmetry classes, respectively. To understand this behavior we   write down $F^{\pm}_{\rm LR}$ from  Eqs.\,(\ref{FanoSS}) at $\varepsilon_{\rm R}=0$ and obtain
\begin{equation}
\label{FanopmSS}
\begin{split}
F^{+}_{\rm LR}(\omega)&=-\frac{1}{\omega}
\left[
\frac{\varepsilon_{\rm L} t}{\omega^{2}-(4t^{2}+\varepsilon_{\rm L}^{2})}\right]\,,\\
F^{-}_{\rm RL}(\omega)&=-
\left[\frac{t}{
\omega^{2}
-(4t^{2}+\varepsilon_{\rm L}^{2})}\right]\,.
\end{split}
\end{equation}
Thus, at vanishing one onsite energy  (e.g., $\varepsilon_{\rm R}=0$), the OTEE class $|F^{+}_{\rm LR}|$ exhibits large values at zero frequency as a function of  $\varepsilon_{\rm L}$ but vanishes at $\varepsilon_{\rm L}=0$, see Fig.\,\ref{Fig5}(a) and the green arrows; see also gray curve in Fig.\,\ref{Fig5}(c). The large zero frequency values of $|F^{+}_{\rm LR}|$ remain robust under variations of $\varepsilon_{\rm L}\neq0$, provided $\varepsilon_{\rm R}=0$; for $\varepsilon_{\rm R}\neq0$, $|F^{+}_{\rm LR}|$ gets large values only near $\omega=0$, see also Fig.\,\ref{Fig5}(b). Thus, when one of the onsite energies is finite and the other vanishes,   $|F^{+}_{\rm LR}|$ develops a divergent profile near zero frequency, see red and green curves in Fig.\,\ref{Fig5}(c) and also gray curve in Fig.\,\ref{Fig5}(d). This divergent profile is similar to what we found for the local pair amplitude discussed in previous subsection   and can be understood to be a consequence of the Majorana nonlocality. This stems from the fact that, given distinct Majorana operators, obeying the self-conjugation Majorana property due to Majorana nonlocality, a pair correlation between two of them still yields a divergent odd-frequency nonlocal pairing, in the same way as it happens for the local pair amplitudes in Eqs.\,(\ref{EQSS}). Thus, having $F_{\rm LR}^+$ as a divergent odd-frequency nonlocal pairing  in the presence of PMMMs can be interpreted  as a measure of Majorana nonlocality.  For finite onsite energies, however, no divergent profile is obtained:  $|F^{+}_{\rm LR}|$ is peaked at   the frequencies of the energy levels occurring away from zero frequency and vanish at $\varepsilon_{\rm R}=\varepsilon_{\rm L}$, see green arrows in  Fig.\,\ref{Fig5}(b) and red curve in Fig.\,\ref{Fig5}(d). Away from this vanishing value but   at distinct onsite energies, we find  $|F^{+}_{\rm LR}|$  to depend linearly  on $\omega$ when approaching zero frequency, see green curve in Fig.\,\ref{Fig5}(d) and third expression in Eqs.\,(\ref{FanoSS}).

In contrast to  the  behaviour of the nonlocal OTEE pairing $F^{+}_{\rm LR}$, 
the nonlocal ETOE pair amplitude  $F^{-}_{\rm LR}$ does not vanish at any onsite energy   and does exhibit any divergent profile at zero frequency even at vanishing onsite energies, see Fig.\,\ref{Fig5}(e,g). This behaviour directly follows Eq.\,(\ref{FanopmSS}), which shows that $F^{-}_{\rm LR}$  only captures the gap edges when either (or both) of the onsite energies vanish. At finite onsite energies,   $|F^{-}_{\rm LR}|$ has a dip near zero frequency, whose minimum value at $\omega=0$ reaches $|F^{-}_{\rm LR}|=|t|/|\varepsilon_{\rm L}\varepsilon_{\rm R}|$, see Eq.\,(\ref{FanoSS}). 

We have therefore obtained that two-site Kitaev chains in the sweet spot $\Delta=t$ exhibit local OTEE pair amplitudes with a divergent odd-frequency dependence around zero frequency, provided one or both onsite energies vanish. Moreover, the nonlocal OTEE pair symmetry class also exhibits a  divergent odd-frequency dependence around zero frequency, as long as either of the onsite energies vanishes and $\varepsilon_{\rm L}\neq\varepsilon_{\rm R}$.   Thus, the divergent odd-frequency profile of the emergent local and nonlocal pair correlations constitute a characteristic of the unconventional superconducting state with PMMMs.   It is worth noting that local and nonlocal pair correlations are expected to play an   important role in local and nonlocal transport \cite{PhysRevB.93.201402,PhysRevB.109.205406}, which happens when attaching leads to the left and right sides of the two-site Kitaev chain studied here. In particular, Andreev reflection and crossed Andreev reflections can be directly determined from the local and nonlocal squared pair  amplitudes \cite{PhysRevB.93.201402,PhysRevB.109.205406}, respectively.  In this regard, having local and  nonlocal pair correlations with distinct functionalities, as we obtain here in the presence of PMMMs, is expected to induce distinct   conductance signatures  that would allow to identify the type of dominant superconducting pairing.

\subsection{Pair amplitudes away from the sweet spot $\Delta\neq t$}
\label{section4d}
\begin{figure}[!t]
\centering
\includegraphics[width=0.49\textwidth]{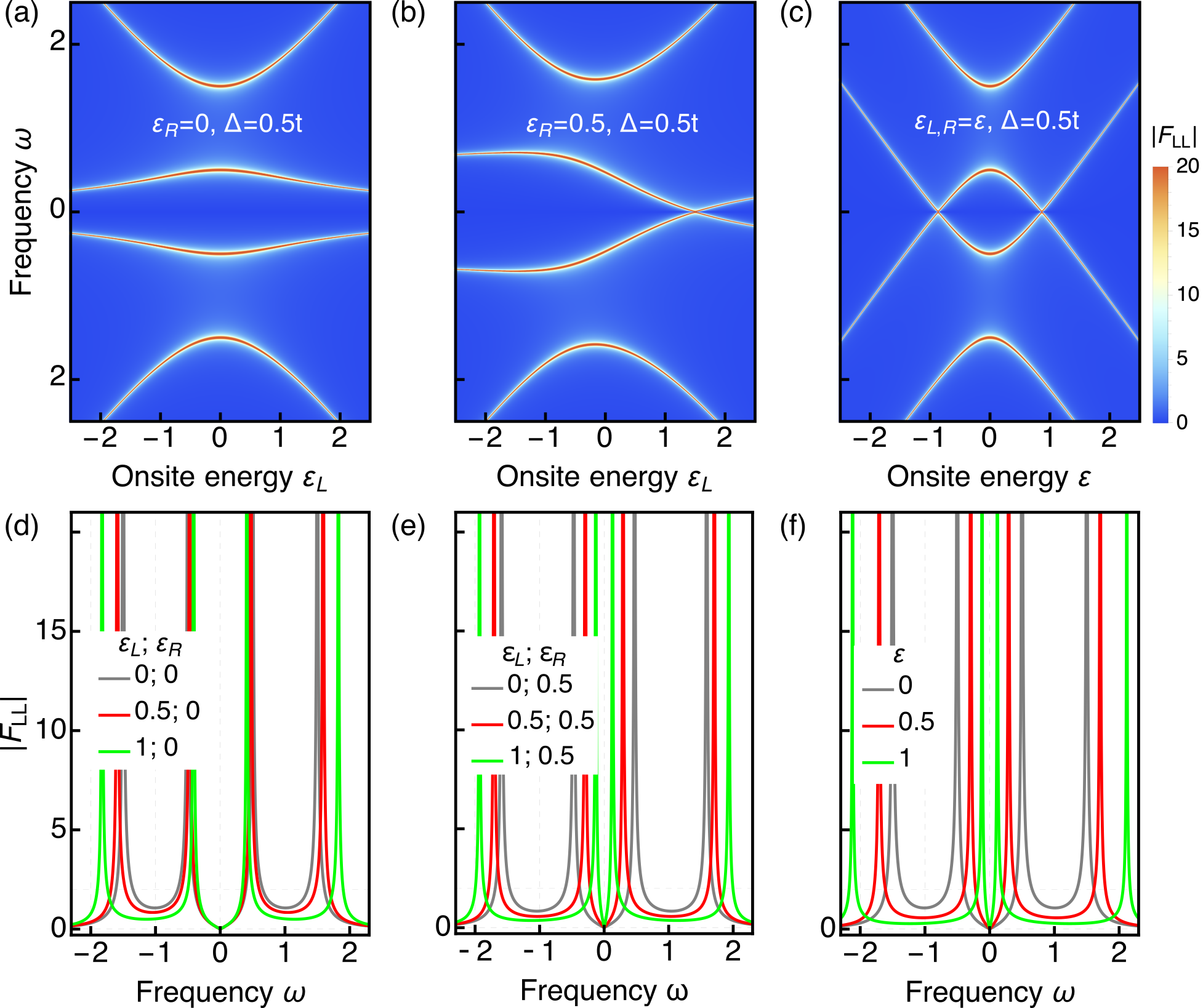}
\caption{(a-c) Absolute value of the local pair amplitude in the left dot $|F_{\rm LL}|$ 
as a function of frequency $\omega$ and onsite energy $\varepsilon_{\rm L}$   at $\Delta=0.5t$ and $\varepsilon_{\rm R}=0$, $0.5$, $\varepsilon_{\rm L}$. (d-f) $|F_{\rm LL}|$ as a function of $\omega$ at distinct $\varepsilon_{\rm L,R}$.}
\label{Fig4} 
\end{figure}

\begin{figure*}[!t]
\centering
\includegraphics[width=0.95\textwidth]{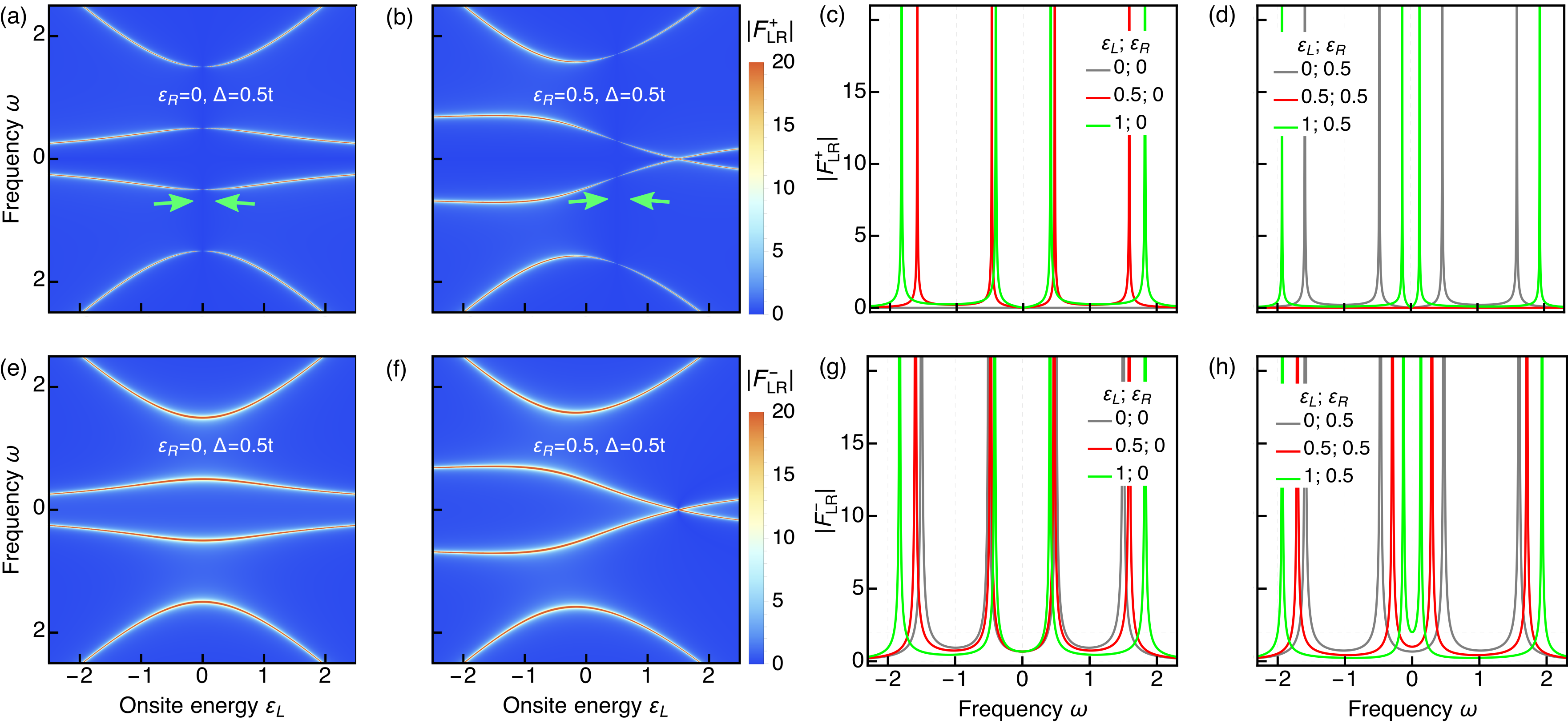}
\caption{(a,b) Absolute value of the nonlocal pair amplitudes $|F_{\rm LR}^{+}|$ as a function of frequency $\omega$ and onsite energy $\varepsilon_{\rm L}$   at $\Delta=0.5t$ and $\varepsilon_{\rm R}=0$, $0.5$. (c,d)  $|F_{\rm LR}^{+}|$ as a function of $\omega$ at distinct values of $\varepsilon_{\rm L,R}$. (e,f,g,h) Same as   (a,b,c,d) but for $|F_{\rm LR}^{-}|$.}
\label{Fig6} 
\end{figure*}

To inspect the behaviour of the emergent pair amplitudes at $\Delta\neq t$, we directly plot Eqs.\,(\ref{Fanomal}) in Figs.\,\ref{Fig4} and Figs.\,\ref{Fig6} for the local and nonlocal components, respectively. In Fig.\,\ref{Fig4}(a-c) and Fig.\,\ref{Fig6}(a,b,e,f)  we show the absolute value of the pair amplitudes $|F_{\rm LL}|$ and $|F^{\pm}_{\rm LR}|$   in the $\omega-\varepsilon_{\rm L}$ plane at  $\varepsilon_{\rm R}=0,0.5,\varepsilon_{\rm L}$. Moreover, Figs.\,\ref{Fig4}(d-f) and Figs.\,\ref{Fig6}(a,b,e,f) we present the frequency dependence of $|F_{\rm LL}|$ and $|F^{\pm}_{\rm LR}|$ at fixed values of $\varepsilon_{\rm L,R}$. The first and general observation in both the local and nonlocal pair amplitudes is that they reveal the energy levels seen in the spectral function in Fig.\,\ref{Fig2}(d-f). There exist, however, slight differences between the local and nonlocal pair amplitudes, which   stems from their particular dependence  on the system parameters since both belong to distinct pair symmetry classes, see Subsec.\,\ref{section4b} and also Sec.\,\ref{section3} and Tab.\,\ref{table:1}.

The local pair amplitudes, which have OTEE symmetry, vanish at zero frequency irrespective of the value of the onsite energies $\varepsilon_{\rm L,R}$, as observed in Fig.\,\ref{Fig4} for $|F_{\rm LL}|$; the same occurs for $|F_{\rm RR}|$. A close inspection reveals that $|F_{\rm LL}|$ approaches zero frequency linearly [Fig.\,\ref{Fig4}(d-f)],  which is  dramatically different to the divergent frequency profile around zero frequency  at the sweet spot $\Delta=t$, see Fig.\,\ref{Fig3}(a,d).  Expanding the first and second expressions  of Eqs.\,(\ref{Fanomal}) at $\omega=0$, and taking only the first order, we obtain $F_{\rm LL/RR}\approx\pm (2t\Delta\omega)/(\Delta^{2}-t^{2}+\varepsilon_{\rm L}\varepsilon_{\rm R})^{2}$, confirming the linear frequency dependence at low frequencies. Moreover, to contrast the divergent and linear frequency dependences of the local pair amplitudes at (away from) the sweet spot $\Delta= t$, it is worth expanding them around $t=\Delta$. Keeping the first order and $\varepsilon_{\rm L,R}=0$, we obtain 
\begin{equation}
\label{FDt}
 F_{\rm LL/RR}(t=\Delta)\approx \pm\frac{2\Delta^{2}}{\omega(\omega^{2}-4\Delta^{2})}\pm\frac{2\omega(t-\Delta)\Delta}{\omega(\omega^{2}-4\Delta^{2})^{2}}\,.
 \end{equation}
 The first term clearly shows the divergent frequency dependence when approaching zero frequency at the sweet spot $t=\Delta$, while the second term reveals that deviations from $t=\Delta$ produces a linear in frequency contribution to the local pair amplitudes.   

For the nonlocal pair amplitudes $|F^{\pm}_{\rm LR}|$,  which posses OTEE and  ETOE symmetries, respectively, we find that the OTEE pair amplitude $|F^{+}_{\rm LR}|$ vanishes either at zero frequency  or when the onsite energies are equal $\varepsilon_{\rm L}=\varepsilon_{\rm R}$. The vanishing   OTEE pair amplitude ($|F^{+}_{\rm LR}|=0$) can be seen in Fig.\,\ref{Fig6}(a,b) and in the gray and red curves of  Fig.\,\ref{Fig6}(c,d); see also green arrows in Fig.\,\ref{Fig6}(a,b) indicating $|F^{+}_{\rm LR}|=0$ at equal onsite energies. To understand how  $|F^{+}_{\rm LR}|$ vanishes, it is useful to expand it at $\omega=0$ and, keeping the first order, we find $F^{+}_{\rm LR}\approx\pm (\omega(\varepsilon_{\rm R}-\varepsilon_{\rm L})\Delta)/(\Delta^{2}-t^{2}+\varepsilon_{\rm L}\varepsilon_{\rm R})^{2}$; this expression also reveals that $|F^{+}_{\rm LR}|$ vanishes in a linear fashion when approaching zero frequency, as indeed seen in Fig.\,\ref{Fig6}(c,d). In contrast to the OTEE pair class, the  ETOE pair amplitude $|F^{-}_{\rm LR}|$ remains finite at zero frequency and when both onsite energies are equal, see Fig.\,\ref{Fig6}(e,f,g,h). It can only vanishe  when $\omega^{2}=\Delta^{2}-t^{2}+\varepsilon_{\rm L}\varepsilon_{\rm R}$, as seen in Eq.\,(\ref{Fanomal}). This is, however, a very stringent condition and the $|F^{-}_{\rm LR}|$ can be thus expected to be in general finite. In fact, by expanding $F^{-}_{\rm LR}$ at zero frequency and keeping the first order, we obtain $F^{-}_{\rm LR}\approx \Delta/(\Delta^{2}-t^{2}+\varepsilon_{\rm L}\varepsilon_{\rm R})$, thus demonstrating that $F^{-}_{\rm LR}$ remains finite even at zero frequency, unlike $|F^{+}_{\rm LR}|$ which vanishes in this case.

\section{Conclusions}
\label{section4}
In conclusion, we have considered few-site Kitaev chains and investigated the emergence of superconducting pair correlations. Under general circumstances, we have shown that distinct pair symmetries are allowed to form in these systems as a result of the multiple existing quantum numbers. In the case of a two-site Kitaev chain, we found    local pairing with an odd-frequency dependence, while nonlocal pair correlations  with both even- and odd-frequency components. While in general the odd-frequency components are linear, we discovered that they acquire a divergent profile around zero frequency   in a fine-tuned sweet spot occurring  when the parent pair potential and electron tunneling are of the same order and at least one onsite energy tuned to zero.  
Since these sweet spot regime corresponds to the phase with poor man's Majorana modes,   which are fully nonlocal, self-conjugate, and charge neutral,  we have interpreted the divergent odd-frequency pairing as a signature of   charge neutrality and spatial nonlocality,  intrinsic to  Majorana quasiparticles. The results presented here can be of use to understand the  nature of superconducting correlations in few-site Kitaev chains.


\section{Acknowledgements}
We thank R. Seoane Souto for insightful discussions. We  acknowledge financial support from the Swedish Research Council  (Vetenskapsr\aa det Grant No.~2021-04121), the Carl Trygger’s Foundation (Grant No. 22: 2093), and the G\"{o}ran Gustafsson Foundation (Grant No. 2216). The computations were enabled by resources provided by the National Academic Infrastructure for Supercomputing in Sweden (NAISS), partially funded by the Swedish Research Council through Grant Agreement No. 2022-06725.

\bibliography{biblio}

\end{document}